\documentclass[aps,twocolumn,showpacs,superscriptaddress,reprint]{revtex4-2}

\usepackage{bm}
\usepackage{graphicx}
\usepackage{amsmath}
\usepackage{amsfonts}
\usepackage{siunitx}
\usepackage{bbold}
\usepackage{comment}
\usepackage{hyperref}

\usepackage{color}

\begin{document}

\title[Anomalous spin textures in a 2D topological superconductor induced by point impurities]%
{Anomalous spin textures in a 2D topological superconductor induced by point impurities}

\author{Dunkan Mart\'{\i}nez}
\thanks{Corresponding author: dunmar01@ucm.es}

\affiliation{GISC, Departamento de F\'{\i}sica de Materiales, Universidad Complutense, E--28040 Madrid, Spain}

\author{\'{A}lvaro D\'{\i}az-Fern\'{a}ndez}
\thanks{Current Address: BBVA Quantum, Calle Azul 4, E--28050 Madrid, Spain}

\affiliation{GISC, Departamento de F\'{\i}sica Aplicada a las Ingenier\'{\i}as Aeron\'{a}utica y Naval, E--28031 Madrid, Spain}

\author{Pedro A. Orellana}

\affiliation{Departamento de F\'{\i}sica, Universidad T\'{e}cnica Federico Santa Mar\'{\i}a, Casilla 110 V, Valpara\'{\i}so, Chile}

\author{Francisco Dom\'{\i}nguez-Adame}

\affiliation{GISC, Departamento de F\'{\i}sica de Materiales, Universidad Complutense, E--28040 Madrid, Spain}

\pacs{
73.63.$-$b;  
73.23.$-$b;  
73.40.$-$c   
}  

\begin{abstract}

Topological superconductors are foreseen as good candidates for the search of Majorana zero modes, where they appear as edge states and can be used for quantum computation. In this context, it becomes necessary to study the robustness and behavior of electron states in topological superconductors when a magnetic or non-magnetic impurity is present. We focus on scattering resonances in the bands and on spin texture to know what the spin behavior of the electrons in the system will be. We find that the scattering resonances appear outside the superconducting gap, thus providing evidence of topological robustness. We also find non-trivial and anisotropic spin textures related to the Dzyaloshinskii-Moriya interaction. The spin textures show a Ruderman–Kittel–Kasuya–Yosida interaction governed by Friedel oscillations. We believe that our results are useful for further studies which consider many-point-impurity scattering or a more structured impurity potential with a finite range.

\end{abstract}

\maketitle

\section{Introduction}

Topological superconductors find a niche of applications in quantum technology since they can host Majorana fermions, at least from a theoretical perspective. Majorana fermions, originally proposed by Majorana once quantum physics was reconciled with the special theory of relativity, are particles that constitute their own antiparticle~\cite{MAJ37}. Although they have not yet been detected as real particles in high-energy physics' experiments, certain low-energy excitations arising in condensed matter physics as edge states of some topological materials have been found to display the theorized characteristics for these fermions~\cite{WIL09}. A Majorana state (also known as Majorana zero mode) can be understood as a fermion with half a degree of freedom~\cite{LEI12} since, in the occupation number formalism, a fermion operator can be rewritten as a sum of two Majorana operators. Because of this, Majorana states appear in pairs. By the same reasoning, if the states are spatially separated, a perturbation that affects one of them will not be able to annihilate it, which gives them great robustness. All this, together with the fact that they present non-abelian statistics, makes them ideal candidates as qubits to achieve noiseless quantum computing~\cite{Read1,Ivanov1,Nayak1,LEI12,SAR15}.

Several proposals have been considered to find signatures of this type of quasi-particles. They involve the use of one-dimensional $p$-type superconductors, in which these particles would appear at their edges~\cite{Kitaev1}, or two-dimensional $p_x + ip_y$ type superconductors, appearing then at the center of vortices~\cite{Volovik1,Read1,Ivanov1}. Both types of superconductors are rare in nature and, therefore, proposals have focused on the use of topological insulators~(TIs) on which a layer of a conventional superconductor is deposited to induce superconductivity by the proximity effect~\cite{FuKane}. In turn, a number of theoretical proposals~\cite{Lutchyn1,Oreg1,FLE10,Prada1,Dunkan1} have been put forward to demonstrate the existence of these quasi-particles. However, its presence in these material systems has not yet been unequivocally determined and this area of condensed matter physics is far from being fully understood.

In order to finally be able to detect Majorana zero modes in an experimental setup, electronic characterization of specific systems and devices is needed. Thus, analyzing how electrons in these materials behave in the presence of impurities becomes essential. In this work, we will work along the lines of Ref.~\cite{Biswas1}, where the effect of single scalar and magnetic impurities at the surface of a TI is analyzed. In our work, we will focus on surface states of a strong TI, such as InSb and HgTe, close to an $s$-wave superconductor. Furthermore, while most previous works deal with zero-range impurity potentials, we introduce an exactly solvable model using a non-local separable pseudo-potential that allows us to address more structured potentials~\cite{Sievert73,Prunele97}. In particular, it is worth mentioning that finite-range pseudo-potentials can nicely reproduce electron interaction with screened, local Coulomb potentials~\cite{Lopez02}. In addition, our approach is particularly useful when extending the study to many impurities by applying the coherent potential approximation, which would allow us to obtain closed expressions for the average density of states, as we showed recently in the case of non-magnetic impurities at the surface of a TI~\cite{Hernando21}.

\section{Model}

In our study, we will consider a two-dimensional TI that supports surface states whose dispersion relation corresponds to Dirac cones. The Hamiltonian of these surface states is that of a single Dirac cone in the pristine material and can be written as~\cite{FuKane}
\begin{equation}
    H_0 = \psi^\dagger(-i\hbar v \boldsymbol{\sigma}\cdot \boldsymbol{\nabla} - \mu)\psi\ .
    \label{eq:TI hamiltonian}
\end{equation}
Here $\psi$ and $\psi^\dagger$ are the electron field operators, including the spin degree of freedom, $\mu$ is the chemical potential, and $v$ is a material characteristic parameter having dimensions of velocity. In order to consider the effect of a single impurity, we will add a non-local separable pseudo-potential to this Hamiltonian~\cite{Sievert73,Prunele97,Lopez02,Lima08,Hernando21}
\begin{equation}
    H_\mathrm{imp} = \psi^\dagger\left|\omega\right> U \left<\omega\right|\psi \ ,
    \label{eq:impurity term}
\end{equation}
where $\omega({\bm r}) = \left< r|\omega \right>$ is referred to as the shape function. The model Hamiltonian then reads $H=H_0+H_\mathrm{imp}$. The intensity of the interaction $U$ can be expressed in terms of an inner product as $U = \vec{\lambda}\cdot \vec{\sigma}$, where $\vec{\lambda} = (\lambda_0,\bm \lambda)$ with $\bm \lambda = (\lambda_x, \lambda_y, \lambda_z)$ being a vector whose components are the coupling constants between the carriers and the impurity. Here $\vec{\sigma}=(\sigma_0,\bm \sigma)$, where $\sigma_0$ is the $2\times 2$ identity matrix and $\bm \sigma=(\sigma_x,\sigma_y,\sigma_z)$ are the spin Pauli matrices.

\begin{figure*}
    \centering
    {\includegraphics[width = \linewidth]{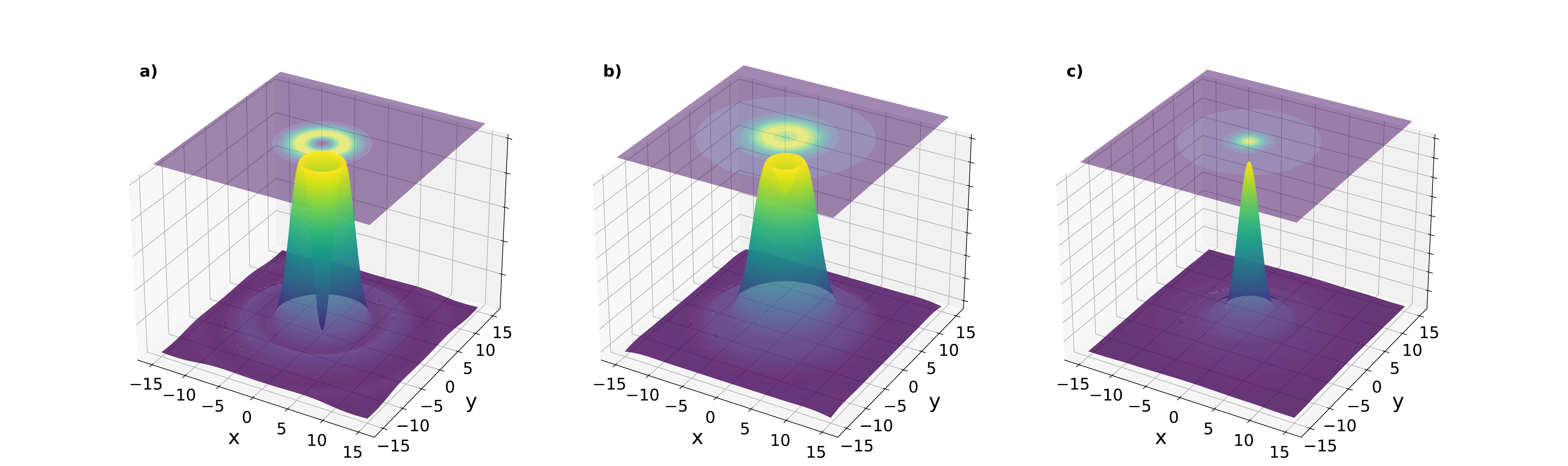}}
    \vspace{-0.55cm}
    \caption{LDOS at resonance energies $\mathrm{Re}(z_R)$ (a)~$0.0067$, (b)~$0.2020$ and (c)~$0.9791$. A height map has been plotted at the top of each figure to highlight the shape of the LDOS.} \label{fig:LDOS}
    \vspace{-0.25cm}
\end{figure*}

On top of this material, a trivial superconductor layer will be deposited in such a way that, due to the proximity effect, the Cooper pairs can tunnel to the surface states from the superconductor. In order to take these processes into account, we have to introduce a new term in the Hamiltonian $V = \Delta\psi_\uparrow^\dagger \psi_\downarrow^\dagger + \text{h.c.}$ where $\Delta = \Delta_0 e^{i\phi}$ is the superconducting gap.

Due to the particle-hole symmetry of the system, we can define the Nambu spinors $\Psi = \begin{pmatrix} \psi_\uparrow, \psi_\downarrow, \psi^\dagger_\uparrow, -\psi_\downarrow^\dagger \end{pmatrix}$, and then use the Bogoliubov-de Gennes Hamiltonian $\mathcal{H}$ where $H = \Psi^\dagger \mathcal{H} \Psi/2$. This can be written as
\begin{multline}
    \mathcal{H} = \tau_z \Big(-i\hbar v \boldsymbol{\sigma} \cdot \boldsymbol{\nabla} - \mu + \left|\omega\right> U \left<\omega\right| \Big) \\ + \Delta_0\left( \tau_x\cos\phi + \tau_y \sin\phi \right)\ ,
    \label{eq: full hamiltonian}
\end{multline}
where $\mu$ is the chemical potential and $\tau_i$ the Pauli matrices associated with particle-hole symmetry. Since the vortex states we are looking for exist for every value of the chemical potential~\cite{FuKane}, we will take $\mu = 0$ for the sake of simplicity of calculations. Furthermore, it is convenient to write the pristine Hamiltonian in momentum space since it will be used in subsequent calculations of the Green's function.
\begin{equation}
    \mathcal{H}_0({\bm k}) = \tau_z\boldsymbol{\sigma} \cdot {\bm k}+ \Delta_0 \left( \tau_x \cos\phi + \tau_y \sin\phi \right)\ .
    \label{eq: pristine hamiltonian}
\end{equation}
Here, energy is measured in units of $\Delta_c$ and momentum in units of $\Delta_c/\hbar v$, with $2\Delta_c$ being the width of the energy region over which the dispersion is linear in momentum.

In order to deal with the impurity term, we will consider the Green's function of the Hamiltonian $\mathcal{H}$ given in~\eqref{eq: full hamiltonian}
\begin{equation}
    G = (z-\mathcal{H)}^{-1} = (1-G_0\mathcal{H}_\mathrm{imp})^{-1}G_0\ ,
\end{equation}
where $G_0$ is the retarded Green's function associated to the pristine Hamiltonian $\mathcal{H}_0$, namely $G_0=1/(z-\mathcal{H}_0)$, and $z = E+i0^+$. We can rewrite this expression as $G = G_0 + G_0 \left|\omega\right> W \left<\omega\right|G_0$, with $W = U/(1-U\left<\omega\right|G_0\left|\omega\right>)$. Here $U$ is expressed in units of $\hbar^2v^2/\Delta_c$. Taking into account the closure relation for the eigenstates of ${\bm r}$, we easily find
\begin{subequations}
\begin{equation}
    G({\bm r}) = G_0({\bm r}) + Q({\bm r})WQ(-{\bm r})\ ,
    \label{eq: G}
\end{equation}
with
\begin{align}
    & G_0({\bm r}) = \frac{1}{(2\pi)^2} \int d^2k\, G_0({\bm k})\ , \label{eq: G0(r)}\\
    & Q({\bm r}) = \frac{1}{2\pi} \int d^2k\, G_0({\bm k})\, \omega({\bm k})\, e^{i{\bm k}\cdot {\bm r}}\ , \label{eq: Q(r)}
\end{align}
and
\begin{equation}
     G_0({\bm k}) = \frac{z + \tau_z \boldsymbol{\sigma} \cdot {\bm k} + \Delta_0(\tau_x \cos\phi + \tau_y \sin\phi)}{z^2-\Delta_0^2-k^2}\ .
\end{equation}
\end{subequations}
Here, $\omega({\bm k})$ is the Fourier transform of the shape function $\omega({\bm r})$. Although the presented results are general for any arbitrary shape function, for illustrative purposes, we will restrict ourselves to a regularized $\delta$-function with Fourier transform $\omega({\bm k})=\omega(k) = \theta(k_c-k)$, with with $k=|{\bm k}|$ and $k_c$ a cut-off momentum which will be chosen to be equal to the momentum at which the energy dispersion ceases to be linear. Defining the matrix $M = z + \Delta_0(\tau_x \cos\phi + \tau_y \sin\phi)$ and considering that the impurity can only be scalar (non-magnetic) or magnetic
\begin{subequations}
\begin{align}
    {}& G_0({\bm r}) = \frac{M}{4\pi} \log \frac{\Delta_0^2 - z^2}{k_c^2 + \Delta_0^2 - z^2}\ , \label{eq: G0 anal}\\
\begin{split}
    {}& Q(\pm {\bm r}) =  M\int_0^{k_c} dk \,\frac{k}{z^2 - \Delta_0^2 - k^2} J_0(kr) \\
    & \quad \quad \pm i\tau_z \sigma_\rho \int_0^{k_c} dk \frac{k^2}{z^2 - \Delta_0^2 - k^2} J_1(kr)\ , \label{eq: Q anal}
\end{split}\\
    {}& W = \frac{1}{1 - \xi^2\lambda^2(z^2-\Delta_0^2)} \bigl[ \tau_z \vec{\lambda} \cdot \vec{\sigma} + \lambda^2 \xi M \bigr]\ . \label{eq: W anal}
\end{align}%
\end{subequations}
where $\xi = \xi(z)= \pi \log\left(\Delta_0^2 - z^2)/(k_c^2 + \Delta_0^2 - z^2\right)$, $J_i(kr)$ are the Bessel functions of the first kind and $\sigma_\rho = \cos\theta \sigma_x + \sin\theta \sigma_y$, being $\theta$ the polar coordinate of the vector ${\bm r}$. Hereafter length will be expressed in units of $1/k_c$. Now we can calculate the Green's function of the perturbed system by means of equation~\eqref{eq: G} and thus obtain the local density of states (LDOS), the spin-polarized local density of states (sLDOS), and the spin textures (ST)~\cite{Biswas1, Alvaro1}
\begin{subequations}
\begin{align}
    \rho({\bm r},E) &= -\frac{1}{\pi} \textrm{Im} \textrm{Tr}[G({\bm r})]\ ,\label{eq: densidad de estados}\\
    \rho^\pm_i({\bm r},E) &= -\frac{1}{\pi} \textrm{Im} \textrm{Tr}\left[ G({\bm r})\left(\frac{\sigma_0 \pm \sigma_i}{2} \right) \right]\ , \label{eq: densidad de estados polarizada}\\
    \boldsymbol{s}({\bm r},E) &= -\frac{1}{\pi} \textrm{Im} \textrm{Tr}\left[ G({\bm r})\frac{\boldsymbol{\sigma}}{2} \right]\ . \label{eq: textura de espín}
\end{align}
\label{eq: todas}%
\end{subequations}
where \textrm{Tr} indicates the trace over the $\tau$ and $\sigma$ degrees of freedom. In view of the definitions~\eqref{eq: todas}, we observe that the contribution of the $\tau$ subspace is solely defined by the Green's function $G({\bm r})$. Moreover, since the trace of the Kronecker product of two matrices is the product of the traces of each one, we can exclude all elements that do not have an identity in the $\tau$ space as they will not contribute to any of the three magnitudes~\eqref{eq: todas}. Therefore, the only relevant contribution is provided by the term
\begin{multline}
        G^u({\bm r}) = \left\{ \frac{z\xi}{4\pi^2} + \lambda \xi z f(z,\lambda) \Bigl[(z^2 + 3\Delta_0^2) F_0^2 + F_1^2\Bigr] \right\} \\
        + i f(z,\lambda) F_0 F_1 z \,\mathbb{1}_2 \otimes [\sigma_\rho, \boldsymbol{\lambda}\cdot\boldsymbol{\sigma}]\ ,
    \label{eq:G util}
\end{multline}
with $F_i = \int_0^{k_c} dk\,k^{i+1} J_{i}(kr)/(z^2 - \Delta_0^2 - k^2) $, $f(z, \lambda) =\lambda/[1 - \xi^2\lambda^2(z^2-\Delta_0^2)]$ and $\mathbb{1}_2$ is the 2$\times$2 identity matrix acting in $\tau$ space. The term on the second line of this equation is traceless, so it will only contribute in the equations \eqref{eq: densidad de estados polarizada} and \eqref{eq: textura de espín} where the Green's function is multiplied by $\sigma_i$. Therefore the LDOS will not depend on the kind of impurity considered. This result can easily be understood as follows. Non-magnetic impurities give rise to resonances at energy $E$ with amplitude $A$, while magnetic impurities give rise to two resonances at $\pm E$ with half the amplitude of the non-magnetic case~\cite{Alvaro1,Biswas1}. Due to the particle-hole symmetry of the superconducting state, both cases behave in the same way, i.e., for non-magnetic impurities, there is an additional resonance at $-E$ energy, while for magnetic impurities, the amplitudes are doubled.

We will focus the study on the resonance energies of the impurity, which can be obtained by the poles of~\eqref{eq: W anal}
\begin{equation}
     \log \left( \frac{\Delta^2 - z_R^2}{k_c^2 + \Delta^2 - z_R^2} \right)^2 (z_R^2 - \Delta^2) = \frac{1}{\pi^2 \lambda^2}\ ,
     \label{eq: Resonancias}
\end{equation}
with $\lambda \equiv |{\vec{\lambda}}|$. This equation cannot be solved analytically. However, we can consider the limit $k_c\to\infty$ to study the behavior of the resonance energy. In such a limit, the resonance energy is located at $z_R^2 \simeq \Delta^2 + 1/(4\pi^2\lambda^2\ln^2k_c)$. Numerically we have found that increasing the value of $k_c$ decreases the energy of the first resonance. Thus, the resonances will always be located outside the gap for any value of the impurity strength, and consequently, there will be no Majorana zero modes in the system.

For the present study, we will consider the $\alpha$-Sn in proximity to a superconducting aluminum layer, so the bandwidth is $\Delta_c = 300\,$meV~\cite{Anh1}, $\hbar v = 100\,$meV\,nm and $\Delta_0 = 0.1\,$meV~\cite{FuKane, Court1}. In addition, as a typical value of the magnetic exchange for topological insulators, we will take $J=300\,\text{meV\,nm}^2$~\cite{PhysRevB1,PhysRevB2}. Consequently, the dimensionless magnitudes turn out to be $\Delta_0 = 3.3\times 10^{-4}$ and $k_c = 1$. As to the coupling between the carriers and the impurity, we will take $\lambda = J/2 = 5$. We can restrict our study to positive energies $\mathrm{Re}(z)\in[0,1]$ thanks to the particle-hole symmetry. Solving the equation \eqref{eq: Resonancias} numerically we find that $\mathrm{Re}(z_R)=0.0067$, $0.2020$ and $0.9791$.

\begin{figure*}[htp]
    \centering
    {\includegraphics[width = 0.9 \linewidth]{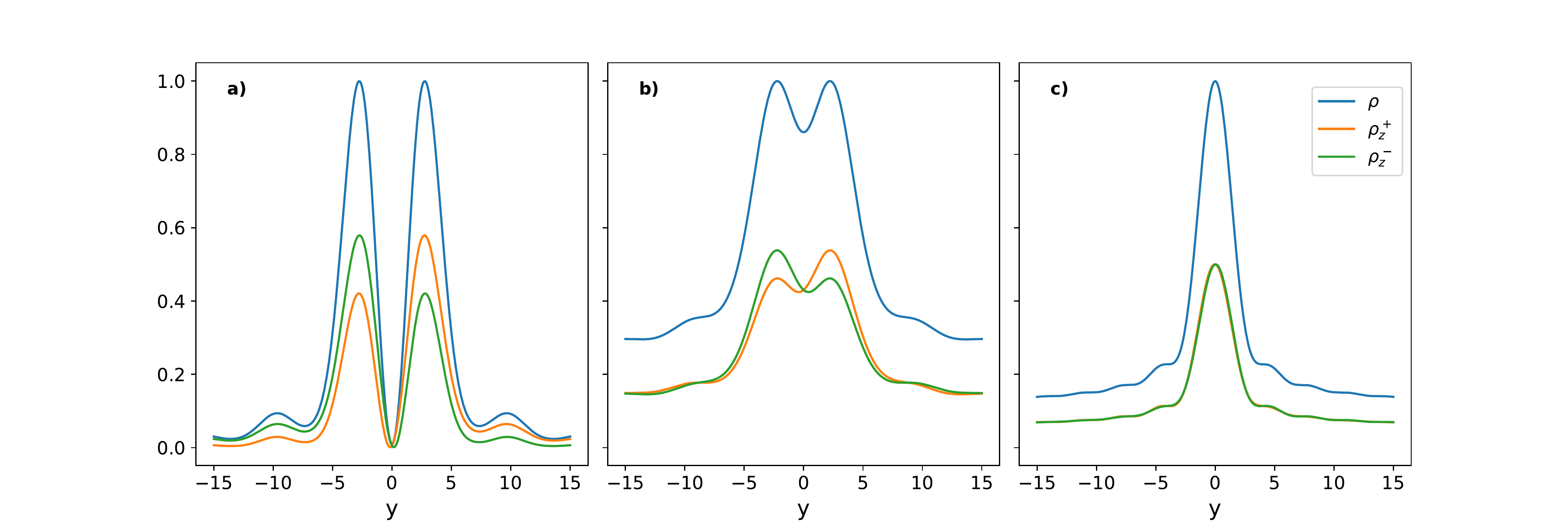}}
    \vspace{-0.25cm}
    \caption{LDOS and sLDOS along the $y$-axis for an impurity with spin oriented parallel to the $x$-axis at energy $\mathrm{Re}(z_R)$ a)  $= 0.0067$, b) $= 0.2020$ and c) $=0.9791$.}
    \label{fig:LDOS polarizadas en espín Sx}
    \vspace{-0.95cm}
\end{figure*}
\begin{figure*}
    \centering
    {\includegraphics[width = 0.99\linewidth]{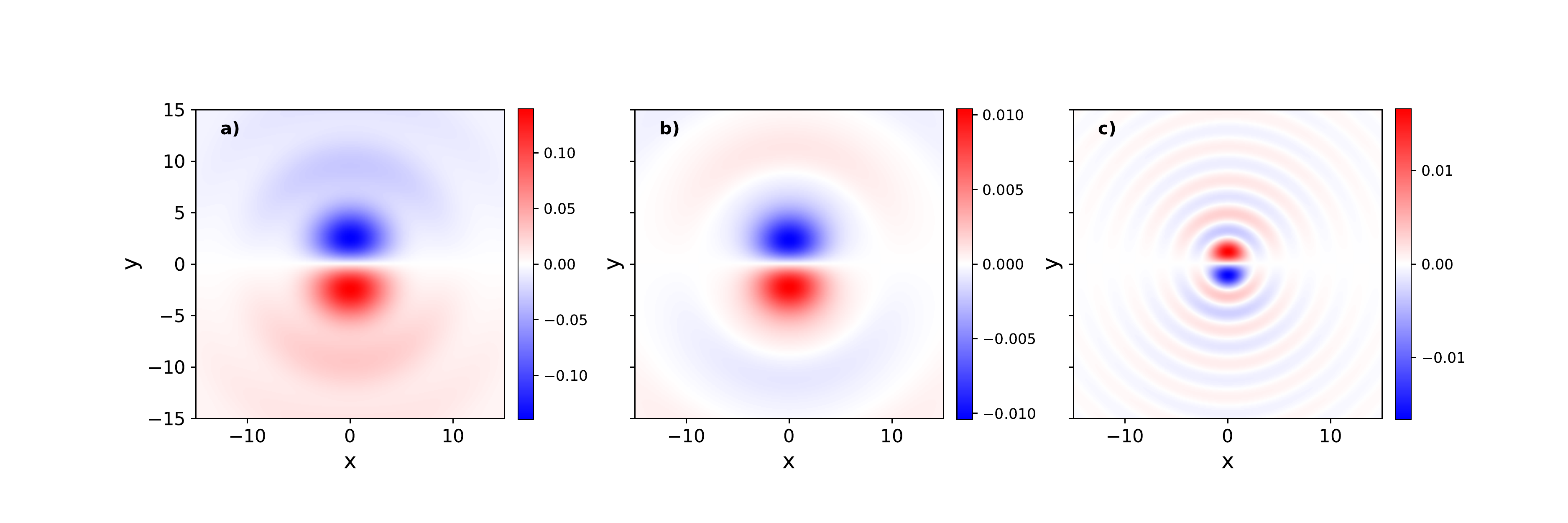}}
    \vspace{-0.8cm}
    \caption{Quasi-normalized ST at energies $\mathrm{Re}(z_R)$ (a)~$=0.0067$, (b)~$= 0.2020$ and (c)~$= 0.9791$ for an impurity spin oriented parallel to the $x$-axis. A color scale has been used for the $z$ component while the $x$ and $y$ components are zero.}
    \label{fig:Texturas de espín Sx}
    \vspace{-0.15cm}
\end{figure*}
\section{Scalar impurity}

Here, we will consider an impurity described by $U = \lambda_0 \sigma_0$. Since we have found that the LDOS does not depend on the kind of impurity introduced into the system but rather on the interaction strength, we will present only LDOS results for this kind of impurity. We can obtain the explicit expression of the LDOS by calculating the trace of $G^u(\bm r)$ given by \eqref{eq:G util}
\begin{multline}
    \rho(\bm{r},z) = - \frac{1}{\pi}\text{Im}\Bigl( \frac{za}{\pi^2} +  4\lambda \xi z f(z,\lambda_0) \\ 
    \times \left[ F_0^2  (z^2 + 3\Delta_0^2) + F_1^2 \right] \Bigr)\ .
    \label{Trace}
\end{multline}
In figure~\ref{fig:LDOS}, the LDOS for the three resonant energies are shown. As the energy approaches $\Delta_c$, we can see that  the resonance becomes narrower, which implies a less effective interaction of the electron with the impurity. On the other hand, we can see that the so-called Friedel oscillations appear, which are more difficult to see as we approach the band edge. It is worth noting that for the lowest resonant energy, the LDOS at ${\bm r} = 0$ is zero. This implies that the presence of the impurity shifts and rearranges the probability density around it, whereas as we move to the band edge, the probability density localizes at the position of the impurity. Furthermore, since the impurity does not generate any interaction with the electron spin, the sLDOS is expected to be half of the LDOS, and therefore the ST is zero.

\begin{figure*}
    \centering
    {\includegraphics[width = 0.9 \linewidth]{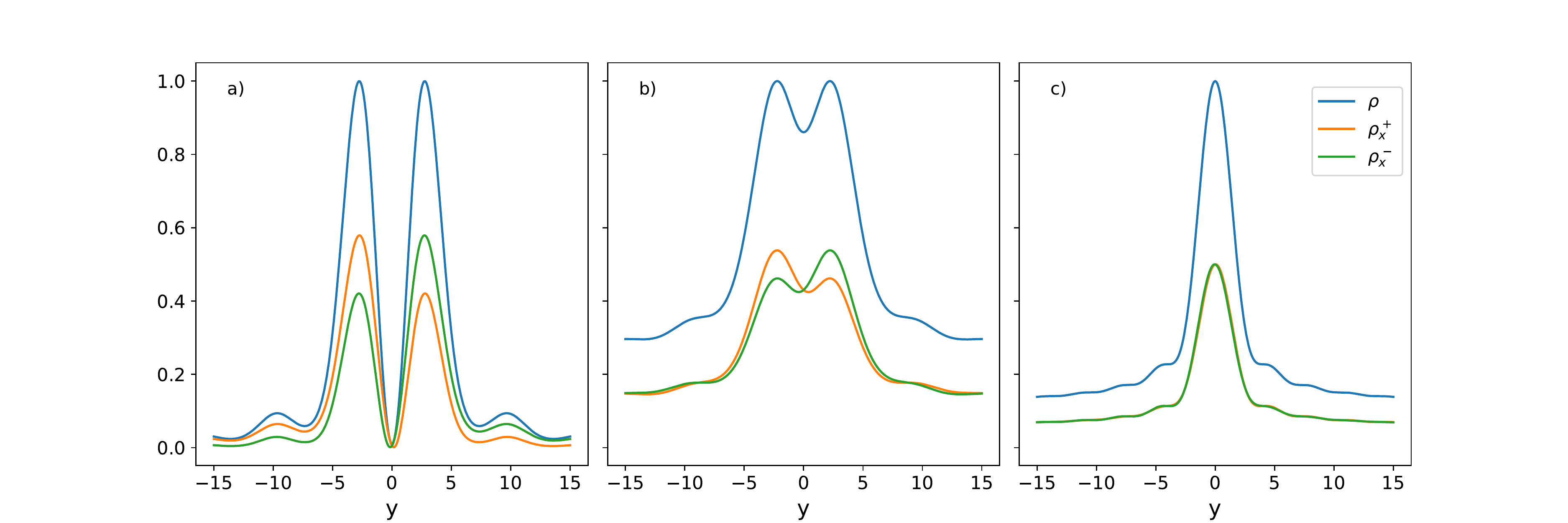}}
    \vspace{-0.2cm}
    \caption{LDOS and sLDOS along the $y$-axis for an impurity with spin oriented parallel to the $z$-axis at energy $\mathrm{Re}(z_R) $ a) $= 0.0067$, b) $=0.2020$ and c) $=0.9791$.}
    \label{fig:LDOS polarizadas en espín Sz}
    \vspace{-0.95cm}
\end{figure*}
\begin{figure*}
    \centering
    {\includegraphics[width = 0.99\linewidth]{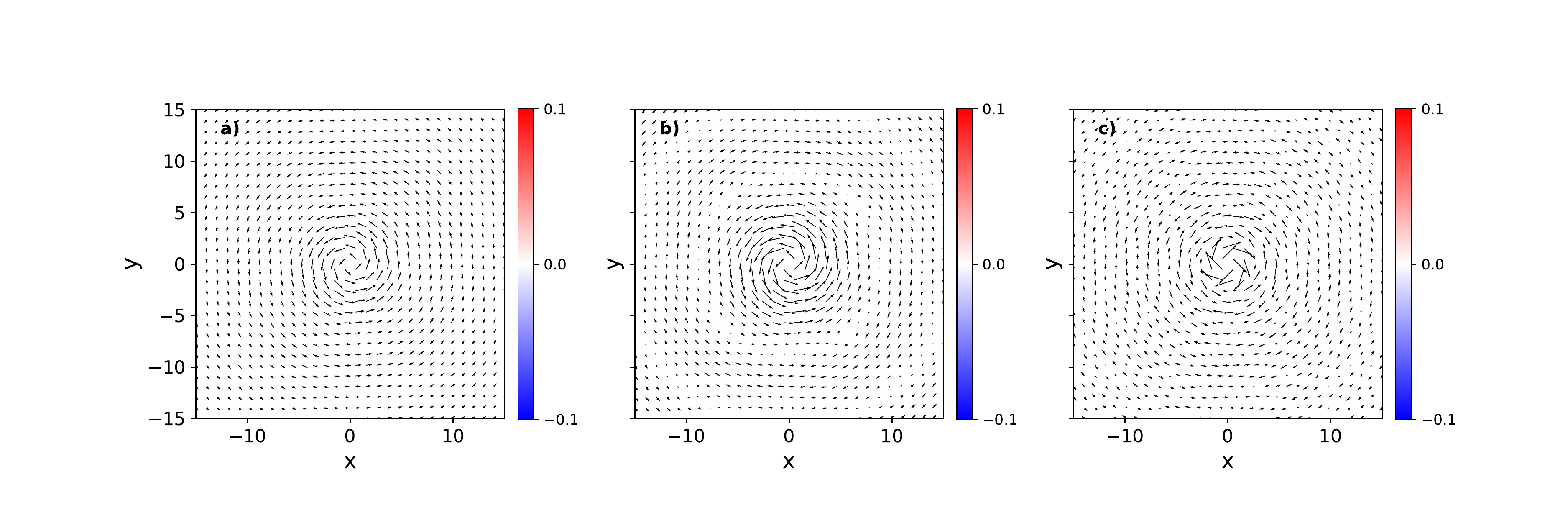}}
    \vspace{-0.8cm}
    \caption{Quasi-normalized ST at energies $\mathrm{Re}(z_R)$ (a)~$=0.0067$ and (b)~$= 0.2020$ for an impurity spin oriented parallel to the $z$-axis. Arrows have been used to represent the $x$ and $y$ directions of the ST while the $z$ direction is zero.}
    \label{fig:Texturas de espín Sz}
    \vspace{-0.15cm}
\end{figure*}
\section{Magnetic impurity}

For this kind of impurity, we will focus on the sLDOS and ST, whose main contribution comes from the second term of equation~\eqref{eq:G util}. This term can be analytically calculated taking into account that $[\bm a \cdot \bm \sigma, \bm b \cdot \bm \sigma] = 2i(\bm a \times \bm b)\cdot \bm \sigma$ for any arbitrary pair of vectors ${\bm a}$ and ${\bm b}$. Considering a magnetic impurity with an arbitrary spin direction $\bm \lambda = (\lambda_x, \lambda_y, \lambda_z)$, equations~\eqref{eq: densidad de estados polarizada} and~\eqref{eq: textura de espín} can be written as
\begin{subequations}
\begin{align}
    \rho_i^\pm (\bm{r},z) &= \frac{\rho(\bm{r},z)}{2} \pm \gamma \hspace{-2pt} \sum_{j=x, y, z}  (\bm \lambda \times \hat{\bm r})_j \delta_{i,j}\ , \label{eq: sLDOS general}\\
    \bm s (\bm{r},z) &=  \gamma \, \bm \lambda \times \hat{\bm r} \ , \label{eq: ST general}
\end{align}
\end{subequations}
where we have defined $\gamma =(4/\pi)\,\textrm{Im}\big[zf(z,\lambda)F_0F_1\big]$ and $\bm \lambda \times \hat{\bm r} = -\lambda_z\sin\theta {\bm e}_x  + \lambda_z\cos\theta {\bm e}_y + (\lambda_x \sin\theta - \lambda_y\cos\theta) {\bm e}_z$. This ST gives rise to a Dzyaloshinskii-Moriya interaction (DMI)~\cite{Biswas1} which is one of the most relevant interactions for specific chiral textures such as magnetic skyrmions. Therefore, our system is prone to have this kind of magnetic texture. However, when we calculate the skyrmion number, we obtain 
\begin{equation}
     \frac{1}{4\pi}\int \bm s \cdot \biggl(\frac{\partial \bm s}{\partial x} \times \frac{\partial \bm s}{\partial y} \biggr) dx\, dy = 0\ .
\end{equation}
Therefore, even though we have DMI, textures such as magnetic skyrmions do not show up.

Finally, we can see that $x$ and $y$ components of both quantities are exactly the same upon changing $\theta\to \theta + \pi/2$ so we will only write $\rho_x^\pm(\bm r, z)$. Since the system is also cylindrically symmetric, we will consider only two types of impurities, one whose spin is oriented in the plane, $\bm{\lambda} = \lambda \bm e_x$, and another whose spin is perpendicular to the plane, $\bm{\lambda} = \lambda \bm e_z$.

\subsection{Impurity with in-plane spin orientation  \label{sec: impureza en x}}

In this situation, the sLDOS can be obtained from equation~\eqref{eq: sLDOS general} taking into account that $\bm \lambda = \lambda_x \bm e_x$. Hence, for this kind of impurity, we get
\begin{subequations}
\begin{align}
    &\rho_x^\pm (\bm{r},z) = \frac{\rho(\bm{r},z)}{2} \label{eq: densidad polarizada en x, Sx}\ , \\
    &\rho_z^\pm (\bm{r},z) = \frac{\rho(\bm{r},z)}{2} \pm \gamma \, \sin\theta \label{eq: densidad polarizada en z, Sx}\ .
\end{align}
    \label{eq: in plane sLDOS}
\end{subequations}

In figure~\ref{fig:LDOS polarizadas en espín Sx} we plot the LDOS and sLDOS for the $z$ spin direction for all energies $\mathrm{Re}(z_R) = 0.0067$, $0.2020$ and $0.9791$. It can be noticed that as the energy increases towards the upper edge of the band, the difference between the spin projection originated by the impurity disappears. This behavior reinforces the claim that the interaction is decreasing. Moreover, we can clearly see Friedel oscillations decreasing in period and amplitude.

The ST can be obtained following equation~\eqref{eq: ST general}. Thus, for this impurity
\begin{equation}
    \bm{s}(\bm{r},z) = \gamma \, \sin\theta  \,\bm e_z\ .
\end{equation}
From figure~\ref{fig:Texturas de espín Sx} we can see that the component along the $z$ direction is asymmetric in the $y$ direction, presenting clear signatures of Friedel oscillations. This could be the reason of the observed variation of the period of the ST as the energy increases. Furthermore, we can notice that as we increase energy, the spin orientation changes and starts to alternate at each oscillation period. In addition, an inversion of the ST near the impurity also occurs at high energies.

\subsection{Impurity with spin oriented perpendicular to the plane\label{sec: impureza en z}}

Following the same steps as in section~\ref{sec: impureza en x}, the sLDOS can be obtained as
\begin{subequations}
\begin{align}
    &\rho_x^\pm(\bm{r},z) = \frac{\rho(\bm{r},z)}{2} \mp \gamma \, \sin\theta\ , \label{eq: densidad polarizada en x Sz}\\
    &\rho_z^\pm(\bm{r},z) = \frac{\rho(\bm{r},z)}{2}\ . \label{eq: densidad polarizada en z Sz}
\end{align}
\end{subequations}
In this case, the ST will vanish in the $z$ direction. In figure~\ref{fig:LDOS polarizadas en espín Sz} we show that the behavior is similar to the one observed in figure~\ref{fig:LDOS polarizadas en espín Sx}. However, this similarity is due to the choice of axes in the figure since now the impurity modifies the spin behavior, both for the $x$ direction and in the $y$ direction, as we will see in the spin textures. Only $\rho_x^\pm(\bm r, z)$ has been shown for all the resonance energies since $\rho_y^\pm(\bm r ,z)$ can be obtained from it and $\rho_z^\pm(\bm r ,z)$ turns out to be zero. The ST can be written as
\begin{equation}
    \bm{s}(\bm{r},z) = -\gamma  (\sin\theta\, \bm e_x - \cos\theta\, \bm e_y)\ .
\end{equation}
From figure~\ref{fig:Texturas de espín Sz} we notice that the STs induced by the impurity are similar to that of a vortex, with a wavelike behavior, as shown in figure~\ref{fig:Texturas de espín Sx}. There is also an alternation of the ST as long as the energy increases and an inversion arises near the impurity.

\section{Conclusions}

Unlike what happens in conventional superconductors~\cite{Matsuura1}, when dealing with topological superconductivity, the introduction of a magnetic impurity does not induce states within the superconducting gap, even when the impurity potential breaks the time-reversal symmetry of the system. 
Furthermore, the LDOS is unaffected by the magnetic or non-magnetic nature of the impurities.
Moreover, the contribution of the topological insulator to equation~\eqref{eq: Q anal} is the only one that introduces a dependence on the type of impurity of the form $[\sigma_\rho, \bm{\sigma}]$ in equation~\eqref{eq:G util}. This dependence makes the magnetic impurities affect only the perpendicular components, i.e., an impurity whose spin is oriented perpendicular to the plane affects only the in-plane  component of the ST and vice-versa. This term is related to the DMI as it represents a strong spin-orbit coupling system with broken symmetry. However, it does not induce topological skyrmions.

Finally, we can highlight a change in the behavior of the spin textures as we increase the energy. Due to the similarity of the obtained spin textures to the results presented by Balatsky \emph{et al}.~\cite{Biswas1} in the context of TIs, we can consider that these changes are related to the Ruderman–Kittel–Kasuya–Yosida interaction so that as we increase the energy and with it the Fermi momentum, the magnetic interaction between two magnetic impurities in this material would change from ferromagnetic to antiferromagnetic, whose behavior is mediated by the spin textures.

\acknowledgments

This work was supported by Spanish Ministry of Science and Innovation (Grant PID2019-106820RB-C21/22) and Recovery, Transformation and Resilience Plan, funded by the European Union - NextGenerationEU (Grant ``Materiales Disruptivos Bidimensionales (2D)” (MAD2D-CM)--UCM5) and FONDECYT grants 1201876 and 1220700. 

\bibliographystyle{apsrev4-2}

\bibliography{references}

\end{document}